\documentstyle[prb,aps,eqsecnum,twocolumn]{revtex}

\begin{document}

\title{A comparison of  FQHE quasi electron trial wave functions on the
sphere}
\author{Uwe Girlich and Meik Hellmund}
\address{Institut f\"ur Theoretische Physik,
Universit\"at Leipzig,
Augustusplatz 10,
04109 Leipzig}

\maketitle

\begin{abstract}
We study Haldane's and Jain's
proposals for the quasiparticle wave function on the sphere.
The expectation values of the energy and the pair angular momenta
distribution are
calculated at filling factor $\frac{1}{3}$ and compared with
the data of an exact numerical diagonalization for up
to $10$ electrons with Coulomb and truncated quasipotential interaction.
\end{abstract}

\narrowtext
\section{Introduction}

The dynamics of interacting planar electrons in the lowest Landau
level (LLL) of a strong magnetic field shows some interesting and not
yet fully understood features
at filling factors $\nu<1$ experimentally observed as the
Fractional Quantum Hall Effect (FQHE)\rlap.\cite{PG90}

Following Haldane's proposal\cite{Hald83,HR85a} we will study
the physics of the FQHE in a spherical geometry.
This gives a clear meaning to the concept of  filling factor
 since the one particle LLL Hilbert
space is finite dimensional. For a sphere\cite{holo}
 penetrated by $2S$ flux
quanta this dimension is $2S+1$ and the filling factor is given by
$\nu=\frac{N-1}{2S}$ for $N$ electrons.

Trial wave functions play an important r\^ole in our attempts to
understand the FQHE. The prime example is the
Laughlin wave function\cite{Lau83b}
which describes very accurately the ground state at filling factors
$\nu=\frac{1}{q}$ with odd integer $q$.

There are two main attempts to understand  other filling factors, the
hierarchy model by  Haldane\cite{Hald83} and Halperin\cite{Halp84}
and the composite fermion (CF) model by Jain\rlap.\cite{Jai89a,Jai90a}
Both describe the reaction of Laughlin's ground state to
the addition or removal of flux quanta  by the
creation of quasiholes or quasiparticles
 and give analytic expressions for the quasiparticle
wavefunctions.

 After
some remarks concerning the structure  of these wavefunctions on the
sphere this paper gives a
 comparison of them with the results of exact numerical
diagonalizations    including up to 10
electrons.
Besides energy expectation values and overlaps we discuss the
distribution
of pair angular momenta as a more demanding test.
Jain's proposal gives better results but contrary to the Laughlin
wavefunction it is not exact even for a electron interaction including
only one quasipotential.

\section{Wave functions on the sphere}

Using  the Wu-Yang\cite{Wu76} gauge for the field of a magnetic
monopole in the center of the sphere (radius $R,$ $e<0,$ $0\leq\theta<\pi$)

\begin{equation}
A_\phi=\frac{\hbar S}{e R}\frac{1-\cos \theta}{\sin \theta},
\end{equation}

a basis
for the $2S+1$ dimensional one particle LLL Hilbert space is
given  by

\begin{equation}\label{basis}
\psi_m(z)={{2S}\choose{S-m}}^\frac{1}{2} z^{S-m},\quad m=-S,-S+1,\dots S.
\end{equation}

Here $z=\tan \frac{\theta}{2}\, e^{-i \phi}$ is the complex stereographic
coordinate.

These wave functions are orthonormal

\begin{equation}
\frac{2S+1}{\pi}\int\frac{d^2z}{(1+|z|^2)^{2S+2}}\; \psi_m^\star(z)\; \psi_n(z)
=\delta_{n,m}
\end{equation}

and related to Haldane's basis by a phase (gauge) factor $e^{iS\phi}$.

In this gauge, the Laughlin function has
up to a normalization factor  $\Pi_i^N { 2S\choose S-m_i}^{1/2}$
the usual shape
$\Psi_L=\Pi_{i<j}^N (z_i-z_j)^q$. In the following we will always
 work with unnormalized  wave functions.

The conserved angular momentum
$\vec l = \vec r \times  (-i\hbar\vec \nabla+e \vec A(\vec r))+\hbar S \vec
r/r$
acts in this basis by

\begin{eqnarray}\label{angular}
l_z&=& S-z\frac{\partial}{\partial z}\nonumber\\
l_+&=&\frac{\partial}{\partial z}\nonumber\\
l_-&=&2Sz-z^2\frac{\partial}{\partial z}.
\end{eqnarray}

The total angular momentum of all electrons
$\vec L=\Sigma\;\vec l_i$ commutes with the two particle  interaction giving a
decomposition of the spectrum into degenerate $SU(2)$ multiplets.
So is   the Laughlin wave function            a  $\vec L=0$ state
with an homogeneous charge distribution and
the quasiparticle appears as a $|\vec L|=N/2$ multiplet.

The first  proposal for such
a quasiparticle multiplet
on the sphere is due to Haldane in the framework of the hierarchy model
and reads in our coordinates\cite{KA92}

\begin{eqnarray}\label{eqhal}
&&\Psi^H(z_0;z_1,\dots,z_N)\nonumber\\
&=&\sum_{m=-N/2}^{m=N/2} z_0^{\frac{N}{2}+m}\Psi^H_m(z_1,\dots,z_N)
\nonumber\\
&=&\prod_{i=1}^{N}
\left(
  q(N-1)
  -z_i\frac{\partial}{\partial z_i}
  +z_0\frac{\partial}{\partial z_i}
\right)
\Psi_L(z_1,\ldots,z_N).\nonumber\\
\end{eqnarray}

One easily checks using eq. \ref{angular} that the $\Psi^H_m$
constitute a $SU(2)$ multiplet of spin $N/2$.

In the CF model, on the other side, a quasiparticle
at $\nu=1/3$
 is described as a state of $N$
composite fermions in a reduced magnetic  flux $2S^\prime=2(S-N+1)$.
Since there is only room for $N-1$ particles in this LLL, one
composite particle has to be in the second Landau level

\begin{equation}
\chi^J_n(z_1,\dots,z_N)=
\left|
\matrix{
z_1^n|z_1|^2 & \cdots & z_N^n|z_N|^2 \cr
1            & \cdots & 1            \cr
z_1          & \cdots & z_N          \cr
\vdots       & \ddots & \vdots        \cr
z_1^{N-2}    & \cdots & z_N^{N-2}
}
\right|
\end{equation}

Now one has  to make an electron wavefunction out of this composite
fermion wave function  according to

\begin{equation}\label{jain1}
\Psi^J_n(z_1,\dots,z_N)= {\cal P}\;\prod_{i<j} (z_i-z_j)^2 \chi^J_n
\end{equation}

where $\cal P$ is the projector to the LLL.
This gives for $n=N/2-1-m=-1,0,\dots,N-2$
only $N$ out of $N+1$ members of an $SU(2)$ multiplet.
Setting $n=N-1$ in eq. \ref{jain1} gives a wrong answer for the last member,
but it can of course
 be obtained by acting with $L_-$
on $\Psi^J_{N-2}.$

Another projection scheme would be\cite{WDJ93}

\begin{equation}\label{jain2}
\Psi^J_n(z_1,\dots,z_N)=\prod_{i<j} (z_i-z_j)\;{\cal P}\;\prod_{i<j}
(z_i-z_j) \chi^J_n
\end{equation}

Remarkably, both schemes give identical wave functions since the
difference between eq. \ref{jain1} and eq. \ref{jain2} includes the factor

\begin{equation}
\sum_i^N \frac{z_i^{n}}{\prod_{j\neq i}(z_j-z_i)} =
\frac{\left|\matrix{
z_1^{n} & \cdots & z_N^{n} \cr
1            & \cdots & 1            \cr
z_1          & \cdots & z_N          \cr
\vdots       & \ddots & \vdots        \cr
z_1^{N-2}    & \cdots & z_N^{N-2}
}
\right|}{
\left|\matrix{
1            & \cdots & 1            \cr
z_1          & \cdots & z_N          \cr
\vdots       & \ddots & \vdots        \cr
z_1^{N-1}    & \cdots & z_N^{N-1}
}
\right|}
\end{equation}

vanishing for $n=0,\dots,N-2$ as well as a factor $n+1$ vanishing
for $n=-1$.

\section{Numerical calculations}

The $2S+1\choose N$--dimensional $N$ electron LLL Hilbert space has a
basis of Slater determinants
$\Psi[m_1,\dots,m_N](z_i)=\det |z_i^{S-m_j}|. $
Since $L_z$ is diagonal  with eigenvalue
$M=\sum m_j,$
it is possible to work in a Hilbert space ${\cal H}_M$ with fixed $M.$
The possible values for $M$ lie between $-M_{max}$ and $+M_{max}$
with $M_{max}=(2S+1-N)N/2.$

 A   generating function
 for its dimension can be found by considering
the grand partition function for a
system having $2S+1$ fermionic energy levels with energy proportional to $m$.
 A Laplace transform projects out the $N$ particle contribution

\begin{eqnarray}
\sum_M \text{dim}({\cal H}_M) x^{M+M_{max}}&=&
\int_0^{2\pi} e^{i N \alpha} \prod_{j=1}^{2S+1}
(1+e^{-i \alpha} x^j) \frac{d\alpha}{2\pi}\nonumber\\
&=&\prod_{j=1}^{N} \frac{1-x^{2S+2-j}}{1-x^j}
\end{eqnarray}

As an illustration, the right hand side evaluates for $N=4$ electrons
at $\nu=1/3,\; 2S=9\; M_{max}=12$ to

\begin{eqnarray}
& &x^{24}+x^{23}+2x^{22}+3x^{21}+5x^{20}+6x^{19}+
9x^{18}+\nonumber\\
&10&x^{17}+13x^{16}+
14x^{15}+16x^{14}+16x^{13}+18x^{12}+\nonumber\\
&16&x^{11}+
16x^{10}+14x^9+
13x^8+10x^7+9x^6+\nonumber\\
&6&x^5+5x^4+
3x^3+2x^2+x+1
\end{eqnarray}
and we can read off from the middle term $x^{12}$ that ${\cal H}_0$ has
dimension 18.

A rotational invariant two particle interaction
in this Slater basis can be expressed as

\begin{equation}\label{hamil}
H=\frac{1}{2}\sum_{m_1,m_2,m_3,m_4} V_{m_1m_2m_3m_4}\;
a_{m_1}^{\dag} a_{m_2}^{\dag} a_{m_3} a_{m_4}.
\end{equation}

where the $a_m^{\dag}$ and $a_m$ create or annihilate an electron in the
state $z^{S-m}.$

 With the help of Clebsch--Gordan
coefficients this is written as sum of contributions from
different pair angular momentum states

\begin{eqnarray}
V_{m_1m_2m_3m_4}&=& \sum_M  \sum_{J=0}^{2S}
\left( {S\atop m_1}{S\atop m_2}\right|\left. J\atop M\right)
\nonumber\\&&\times
\left( {S\atop m_3}{S\atop m_4}\right|\left. J\atop M\right)
\;V^{(S)}_{2S-J}.
\end{eqnarray}

Due to Fermi statistics only odd quasipotential coefficients\cite{Hald83}
$V_i, i=1,3,\dots$  contribute to eq. \ref{hamil}.
For the case of  Coulomb interaction  proportional to $(\text{chord
distance})^{-1}$ the  $V_i$ are calculated by
Fano et al\rlap.\cite{FOC86}

 We consider  also the
truncated quasipotential model,
 where two particles repel each other only if they are in
a state of maximal relative angular momentum, i.~e. only $V_1$ is
nonvanishing.

The lowest eigenvalue and the corresponding eigenvector of the
Hamiltonian eq.~\ref{hamil}
were calculated by an iterative Lanczos procedure\rlap.\cite{CULL85}
In order to handle the very large  matrices
(of, e.~g., dimension $165\,821^2$ for $N=10$) a
sophisticated algorithm for indexing and storing was used.

To calculate
the energy  (and other) expectation values of the trial
 wave functions they are  expanded in Slater states
in order to have  them in
the same form as the exact eigenvectors\rlap.\cite{apel}

\section{Results}

Fig. \ref{figEn} shows the
finite size dependence of the
energies for $V_1$ interaction. The
data are   fitted\cite{DM89} by a quadratic
polynomial in $N^{-1}.$ Obviously, Jain's wavefunction works better.
In the $N\rightarrow\infty$ limit its energy is about $7\%$ too high.
This has to be compared with an $28\%$ error in the  hierarchy model.
 The results are qualitatively similar for Coulomb interaction and are
confirmed by a calculation of the overlaps of the different wave
functions, shown for $N=10$ electrons in Table \ref{tabOver}.
The superiority  of the CF quasiparticle wave function
in the reproduction of finite size calculations
has also been found in
the disk geometry\rlap.\cite{KA93a}

 To give an optical impression of a quasiparticle, Fig. \ref{figCharge}
shows the
charge distribution of the $L_z=\frac{N}{2}$ member of the
quasiparticle multiplet. In the sense of eq. (\ref{eqhal}) this can be
interpreted as a quasiparticle sitting at the south pole $z\rightarrow\infty.$
As already observed in the disk geometry\rlap,\cite{DM89} the
quasiparticle excess charge is concentrated on a ring of roughly the size of
a magnetic length  ($R/\sqrt{S}$ in our units).
Higher quasipotentials introduce
more inhomogeneities outside the excitation as
 the case of Coulomb interaction (dashed line) shows.

As emphasized by Gros and MacDonald\rlap,\cite{GM90}
a crucial r\^ole in the dynamics of the FQHE is played by the
distribution of pair
angular momenta.
The Laughlin state  is the only state in the $\nu=1/3$ Hilbert space
with vanishing contribution for the highest possible pair angular
momentum $2S-1.$ For higher filling factors (i.~e. in the presence of
quasiparticles) no such state exists in the Hilbert space.
But due to the $V_1$ interaction, the one quasiparticle
ground state is still a state
with very small probability of finding an electron pair with angular
momentum $2S-1.$ These probabilities are shown for a $N=10$ electron
system in Fig.~\ref{figAng} normalized to sum up to the number of pairs
$N(N-1)/2=45.$
One sees clearly, how the interaction nearly empties the
pair states coupling to $V_1$ resulting in a very high probability
for pairs in the next-to-highest angular momentum state.
This  picture is not very much changed in the Coulomb case, an
astonishing result facing the fact that the $V_3$ interaction should
suppress the amplitude of the next-to-highest angular momentum state.
The smallness of this suppression confirms the point of view that the
$V_1$ hard core model includes the essential physics of the FQHE and
higher quasipotentials just make small qualitative  changes.
It seems that even in the hard core model the amplitudes for higher
angular momentum pairs have already nearly reached their maximum for this
Hilbert space.
It would be nice to have a theoretical insight in the
occurrence of these
filling factor dependent maximal and
minimal probabilities of finding electron pairs in states of
some relative angular momentum.

Remarkably, both the hierarchy and the CF proposal reproduce very well
the exact distributions.
This supports the conclusion that both of them capture in their
analytic expressions some of the essential physics of the FQHE.
The ground state energies show, however, a
much better behaviour for the CF wavefunction especially in the large
$N$ extrapolation.

\acknowledgements
We are grateful for many stimulating and valuable discussions with
W. Apel, M. Kasner and W. Weller.

\newpage
%\bibliographystyle{prsty}
%\bibliography{fq,numerik,twf}

\begin{thebibliography}{10}
\bibitem{holo}
Since the Hilbert space is the space of holomorphic sections of a line bundle
  with Chern number $2S,$ we get from the Riemann--Roch theorem a dimension
  $2S+1-g$ for a genus $g$ surface if $S>g-1$.
\end{thebibliography}

\newpage
\begin{figure}
\caption{Quasiparticle energy for $V_1$ interaction vs. $N^{-1}.$}
\label{figEn}
\end{figure}

\begin{figure}
\caption{Charge density distribution $\rho(\theta)$  of 10 electrons for
one quasiparticle at the south pole normalized to
$\int \rho(\theta) \; d\cos \theta = 10 |e|.$}
\label{figCharge}
\end{figure}
\begin{figure}
\caption{Pair angular momenta distribution.}
\label{figAng}
\end{figure}

%\newpage

\begin{table}
\caption{Quasiparticle wave function overlaps for $N=10.$}
\begin{tabular}{lccc}
                 & exact ($V_1$) & Jain & Haldane \\
\hline
exact (Coulomb)  & 0.987584      & 0.985416     & 0.972998     \\
exact ($V_1$)    &               & 0.988918     & 0.968469     \\
Jain             &               &              & 0.993149     \\
\end{tabular}
\label{tabOver}
\end{table}

\end{document}